\documentclass[prl,showpacs,twocolumn]{revtex4}
\usepackage{amsmath}
\usepackage{epsfig}          
\usepackage{longtable}
\begin{document}

\title{Bimodality in the transverse fluctuations of a grafted semiflexible polymer and the diffusion-convection  analogue: an effective-medium approach }
\author{P. Benetatos$^1$, T. Munk$^2$,  and E. Frey$^2$}
\affiliation{$^1$Hahn-Meitner-Institut, Abteilung Theoretische Physik,
Glienicker Stra{\ss}e 100, D-14109 Berlin, Germany\\
$^2$Arnold Sommerfeld Zentrum f\"ur Theoretische Physik, Department f\"ur Physik, Ludwig-Maximilians-Universit\"at M\"unchen, Theresienstra{\ss}e 37,
D-80333 M\"unchen, Germany}


\date{\today}

\begin{abstract}

Recent Monte Carlo simulations of a grafted semiflexible polymer in $1+1$ dimensions have revealed a pronounced bimodal structure in the probability distribution of the transverse (bending) fluctuations of the free end, when the total contour length is of the order of the persistence length [G. Lattanzi {\it et al.}, Phys. Rev E ${\bf 69}$, 021801 (2004)]. In this paper, we show that the emergence of bimodality is related to a similar behavior observed when a random walker is driven in the transverse direction by a certain type of shear flow. We adapt an effective-medium argument, which was first introduced in the context of the sheared random-walk problem [E. Ben-Naim {\it et al.}, Phys. Rev. A ${\bf 45}$, 7207 (1992)], in order to obtain a simple analytic approximation of the probability distribution of the free-end fluctuations. We show that this approximation captures the bimodality and most of the qualitative features of the free-end fluctuations. We also predict that relaxing the local inextensibility constraint of the wormlike chain could lead to the disappearence of bimodality.

\end{abstract}

\maketitle

Semiflexible polymers have been the on focus of intense theoretical and experimental research activity in recent years. The main reasons are their relevance to biology, as many biologically important macromolecules fall in this category (including the building blocks of the cytoskeleton) \cite{Howard,Nelson}, and inherent challenges in their theoretical study. A widely used minimal theoretical model of semiflexible polymers is the wormlike chain, a locally inextensible, fluctuating line with bending stiffness \cite{STY}. The main parameter in the description of a wormlike chain is the persistence length, $L_p$. It is defined as the correlation length of the tangent unit vector along the polymer contour, and it is proportional to the bending stiffnes, $\kappa$: $L_p=2\kappa/[k_B T(d-1)]$, where $T$ is the temperature, and $d$ is the dimensionality of the embedding space. The flexible limit of the wormlike chain, $L \gg L_p$ ($L$ being the total contour length), and to a lesser extent the weakly-bending limit, $L \ll L_p$, have been well studied theoretically. In both of these extreme cases, the wormlike chain exhibits scaling behavior. In the intermediate regime where the total contour length is of the order of the persistence length, however, scaling ceases to exist and intriguing new phenomena emerge. One of the most striking features in the fluctuations of a polymer in this intermediate regime was revealed in recent Monte Carlo simulations of a grafted wormlike chain in a two-dimensional embedding space \cite{Lat}. A grafted polymer has both the position {\it and} the orientation at one end fixed. If we look at the probability distribution of the transverse (bending) displacement of the free end, it starts as a delta function in the rigid-rod limit, which develops into a Gaussian in the weakly-bending regime, and it becomes Gaussian again (of a different type) in the flexible regime. Surprisingly, in the intermediate region, the probability distribution is not a smooth interpolation between the two Gaussian limits, but it displays a pronounced bimodality. One can safely claim that this bimodality is the hallmark of semiflexibility.

We should point out that the emergence of bimodality (or multimodality) from an initially unimodal probability distribution upon the variation of a time-like variable is not peculiar to the semiflexible polymers. It has been shown that L\'evy flights in steeper than harmonic potentials exhibit a critical time beyond which an initially unimodal distribution evolves into a bimodal terminal one \cite{Metz1}. Depending on the steepness of the confining potential, a trimodal transient may exist \cite{Metz2}. Another manifestation of bimodality which is formally closer to that in semiflexible polymers occurs in diffusion-convection problems. If a neutral Brownian particle is carried by a power-law shear flow in the transverse direction, then the probability distribution of transverse displacements gives rise to a terminal bimodality, depending on the exponent of the flow profile \cite{Red2}. A similar bimodality occurs when the convective flow is odd and random (for a single realization of the randomness) \cite{Red1}. The most important difference between the above mentioned cases and the probability distribution of transverse fluctuations in the grafted semiflexible polymer is that, in the former, bimodality is terminal, whereas in the latter it is transient.

If $G[${\bf r}$(L), \theta(L)]$ is the probability distribution function for a wormlike chain having its free end at point ${\bf r}(L)=[x(L),y(L)]$ with a tangent vector making an angle $\theta(L)$ with respect to the clamping direction ($x$-axis), given that the other end is at the origin of the coordinate system, it obeys the equation \cite{ftnt}:
\begin{eqnarray}
\label{wlceq1}
& & \Big[\frac{\partial}{\partial L}+\cos\theta\frac{\partial}{\partial x} +\sin\theta \frac{\partial}{\partial y}- \frac{1}{L_p}\frac{\partial^2}{\partial \theta^2}\Big]\nonumber\\
& & \times G[{\bf r}(L), \theta(L)]=0.
\end{eqnarray}
If we integrate out the spatial degrees of freedom ($x$ and $y$), the resulting diffusion equation for the angle is simply the differential equation associated with a path integral of the Boltzmann weight of the bending energy \cite{STY}. The ``convective'' terms express the local inextensibility of the wormlike chain \cite{Daniels}. 

If we integrate out the longitudinal fluctuations, $x(L)$, we obtain:
\begin{eqnarray}
\label{wlceq2}
\Big[\frac{\partial}{\partial L}+\sin\theta \frac{\partial}{\partial y}- \frac{1}{L_p}\frac{\partial^2}{\partial \theta^2}\Big]G_y[y(L), \theta(L)]=0.
\end{eqnarray}
Integrating out the angle in $G_y[y(L), \theta(L)]$ would give us the desired distribution which was observed in the simulation of Ref. \cite{Lat}. Eq. (\ref{wlceq2}) can be simplified and solved analytically (using Fourier transformations) in the weakly bending limit ($L\ll L_p$), where $\theta \ll 1$ and $\sin \theta \approx \theta$ \cite{Ben}:  
\begin{equation}
\label{wb}
G_y(y,\theta)=\frac{L_p\sqrt{3}}{\pi L^2}\exp\Big\{-\frac{L_p\theta^2}{2L}\Big\}\exp\Big\{-\frac{6L_p[y-\langle v \rangle L]^2}{L^3}\Big\}\;,
\end{equation}
where $\langle v \rangle = \theta/2$. We have introduced the extra symbol $v$ because we want to invoke the analogy with the diffusion-convection problem of Ref. \cite{Red2}. According to this analogy, Eq. (\ref{wlceq2}) describes the motion of a particle in a two-dimensionsional space ($\theta$ and $y$), with $L$ becoming the time variable: it diffuses in the $\theta$-direction and it gets carried by a sinusoidal flow in the $y$-direction. The flow is ${\bf v}(\theta,y)=(\sin\theta)\hat{\bf y}$. Note that, in contrast to the flows considered is Refs. \cite{Red1} and \cite{Red2}, this flow has a periodic profile. The weakly-bending limit of the wormlike chain corresponds to the case of linear shear flow. In that case,  $\langle v \rangle = \theta/2$ is the average velocity of the $y$-coordinate in the interval  $[0,\theta]$.

Before we implement the effective-medium approximation, we should mention that attempts to perturbatively capture the onset of bimodality (from the stiff limit) have failed. Specifically, we had expanded $\sin\theta$ in Eq. (\ref{wlceq2}) up to the quintic term, and we had treated the nonlinear terms to lowest order (one-loop) in perturbation theory. We had also used the analytic expressions for the variance and the kurtosis of the exact distribution in an Edgeworth expansion \cite{Feller} about the Gaussian, again without success.


\begin{figure}
\begin{center}
\leavevmode
\hbox{%
\epsfxsize=2.5in
\epsffile{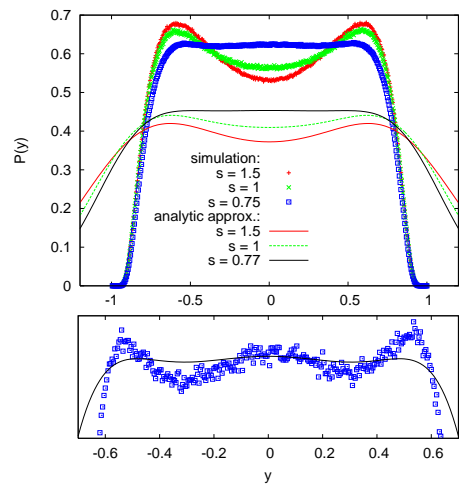}}
\end{center}
\caption{\label{fig1} The distribution funcion of bending displacements, $P(y)$, from the analytic approximation and from the simulation, approaching bimodality from the the stiff limit. $y$ has been rescaled to $y/L$ and $s\equiv L/L_p$. The lower panel shows the onset of multimodality with the three peaks.}
\end{figure}
\begin{figure}
\begin{center}
\leavevmode
\hbox{%
\epsfxsize=2.5in
\epsffile{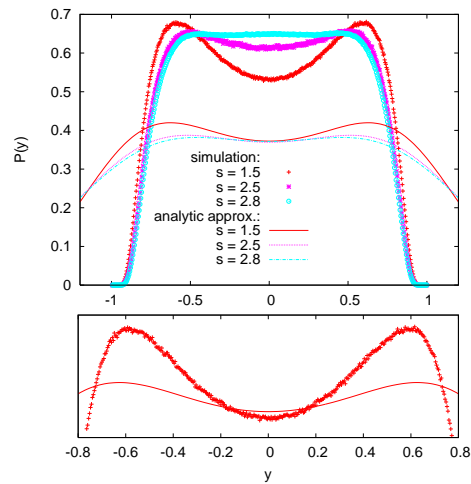}}
\end{center}
\caption{\label{fig2} The distribution funcion of bending displacements, $P(y)$, from the analytic approximation and from the simulation, as the polymer approaches the flexible region. The lower panel shows a detail from the region where bimodality is most pronounced.}
\end{figure}


In the effective-medium approach, following Refs. \cite{Red1} and \cite{Red2}, we hypothesize that the probability distribution of the transverse displacement and the slope of the free end maintains the form of Eq. (\ref{wb}) for arbitrary bending stiffness (arbitrary $L/L_p$), but with the average ``velocity'' $\langle v \rangle$ now being replaced by an ``effective velocity,'' namely $\sin\theta$. In this approximation, the probability distribution of transverse fluctuations reads:
\begin{eqnarray}
\label{pofy}
& &P(y)=\frac{L_p\sqrt{3}}{\pi L^2}\int_{-\infty}^{\infty}d \theta \exp\Big\{-\frac{L_p\theta^2}{2L}\Big\}\nonumber\\
& &\times\exp\Big\{-\frac{6L_p[y-(\sin\theta) L]^2}{L^3}\Big\}\;.
\end{eqnarray}
The integral can be evaluated numerically, and the result compared with the Monte Carlo simulation data is illustrated in Figs. \ref{fig1} and \ref{fig2}. In the weakly-bending limit, the analytic approximation gives a narrow Gaussian with a width $\langle y^2 \rangle=(13/6)L^3/L_p$ (there is a discrepancy by a numerical prefactor of about $0.31$ with the exact result). As $L/L_p$ increases, the distribution flattens and it develops three peaks at $L/L_p\approx 0.77$ in the analytic approximation and at $L/L_p\approx 0.75$ in the simulation. Increasing the flexibility of the chain, the central peak gets suppresed, and a clear double-peak structure emerges, which is most pronounced around $L/L_p\approx 1.5$ both in the analytic approximation and in the simulation. The results for the position of the peaks from the two approaches are quite close. Eventually, the bimodality gets suppressed and a unimodal distribution reemerges in the flexible regime. This happens at $L/L_p\approx 2.8$ in the simulation and at $L/L_p\approx 3.8$ in the analytic approximation. The reentry into unimodality is accompanied by a transient trimodality in the simulation whereas this trimodality does not appear in the analytic approximation. The major drawback of the effective-medium approximation is that it fails to yield the right scaling for the width of the Gaussian distribution in the flexible regime. It reproduces $\sqrt{\langle y^2 \rangle} \sim L^{3/2}$ of the weakly bending limit, instead of $\sqrt{\langle y^2 \rangle} \sim L^{1/2}$ of the Gaussian chain. It also has Gaussian tails which spread into a region of extreme dispacements which is forbidden by the inextensibility constraint of the wormlike chain.

\begin{figure}
\begin{center}
\leavevmode
\hbox{%
\epsfxsize=2.5in
\epsffile{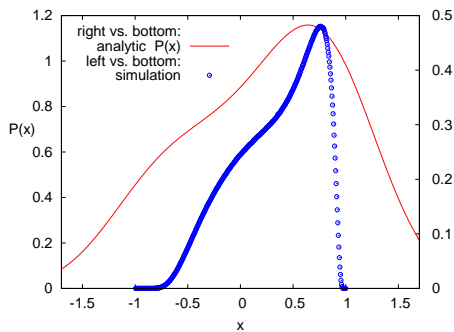}}
\end{center}
\caption{\label{fig3}The distribution funcion of longitudinal displacements, $P(x)$, from the analytic approximation and from the simulation, for $L/L_p=2.8$, where it exhibits two convex bumps. $x$ has been rescaled to $x/L$.}
\end{figure}

We now consider the probability distribution of the longitudinal displacements of the free end. If we integrate out the transverse displacements, $y(L)$, in Eq. (\ref{wlceq1}), we obtain:
\begin{equation}
\label{wlceq3}
\Big[\frac{\partial}{\partial L}+\cos\theta \frac{\partial}{\partial x}- \frac{1}{L_p}\frac{\partial^2}{\partial \theta^2}\Big]G_x[x(L), \theta(L)]=0.
\end{equation}
Applying the effective-velocity approximation to this equation, and integrating out the orientational degree of freedom, we obtain:
\begin{eqnarray}
\label{pofx}
& &P(x)=\frac{L_p\sqrt{3}}{\pi L^2}\int_{-\infty}^{\infty}d \theta \exp\Big\{-\frac{L_p\theta^2}{2L}\Big\}\nonumber\\
& &\times\exp\Big\{-\frac{6L_p[x-(\cos\theta) L]^2}{L^3}\Big\}\;.
\end{eqnarray}
Plotting $P(x)$ for various values of $L/L_p$, we obain a distribution which, apart from the Gaussian-fat tails, qualitatively agrees with the results from the simulation. Specifically, the value of $x$ which corresponds to the peak of the distribution is pretty close to that obtained from the simulation over a wide range of stiffness (Table I). The agreement is better in the weakly-bending region. A remarkable feature of this distribution is that, although it always has only one peak, for $1.8\lesssim L/L_p \lesssim 3.4$, it exhibits two convex bumps. This structure was {\it predicted} by the analytic approximation and was subsequently confirmed by the simulation. The presence of the two bumps for $L/L_p=2.8$ is shown in Fig. \ref{fig3}. Note that this structure is somehow reminiscent of the double-peaked structure in the spatial density function of a (free) wormlike chain which was observed in Monte Carlo simulations by Dhar {\it et al.} in Ref. \cite{Dhar}, in an overlapping region of $L/L_p$. Using Eq. (\ref{pofx}) to calculate the average longitudinal position of the free end, $\langle x \rangle$, we obtain a particularly simple expression:
\begin{equation}
\label{avex}
\langle x \rangle = L \exp\Big\{-\frac{L}{2L_p}\Big\}\;.
\end{equation}
In Table II, we show how this approximate result compares with the value of $\langle x \rangle$ obtained from the simulation. The two results are very close in the weakly-bending regime, but they diverge very quickly as the wormlike chain enters the flexible regime.

\begin{widetext}


\begin{table}[htbp]
\begin{tabular}{c|cccccccccc}
$L/L_p$&0.1&0.2&0.3&0.4&0.5&0.75&1&1.5&2.5&2.8\\\hline\hline
simulation
$x_\mathrm{max}/L$&0.992(1)&0.984(1)&0.977(1)&0.968(1)&0.960(1)&0.940(2)&0.922(2)&0.878(2)&0
.798(4)&0.762(4)\\
analytic
$x_\mathrm{max}/L$&0.967&0.943&0.925&0.908&0.893&0.860&0.827&0.770&0.684&0.653\\\hline
$\delta$(\%)&2.5(1)&4.2(2)&5.3(2)&6.2(2)&7.0(2)&8.5(3)&10.3(3)&12.3(3)&14.3(6)&14.3(6)\\
\end{tabular}
\caption{Longitudinal position of the free end at the peak of the probability distribution, from the simulation, from the analytic approximation, and their
relative difference, for various values of the flexibility, $L/L_p$.}
\label{tab:1}
\end{table}
\vspace{-5mm}
\begin{table}[!ht]
\begin{minipage}[t]{.29\textwidth}
\vspace{-5mm}\caption{\label{tab:2}Average longitudinalposition of the free end, from the simulation, from the analytic approximation, and their relative
difference, for various values of the flexibility, $L/L_p$.}
\end{minipage}
  \begin{minipage}[t]{.7\textwidth}
\begin{tabular}[t]{c|cccccccccc}
$L/L_p$&0.1&0.2&0.3&0.4&0.5&0.75&1&1.5&2.5&2.8\\\hline\hline
simulation $\langle
x/L\rangle$&0.953&0.910&0.869&0.829&0.792&0.711&0.639&0.525&0.373&0.340\\
analytic $\langle
x/L\rangle$&0.951&0.905&0.861&0.819&0.779&0.687&0.607&0.472&0.287&0.247\\\hline
$\delta$(\%)&0.2&0.5&0.9&1.0&1.6&3.4&5.0&10.1&23.1&27.4\\
\end{tabular}
\end{minipage}
\end{table}


\end{widetext}


\begin{figure}
\begin{center}
\leavevmode
\hbox{%
\epsfxsize=2.5in
\epsffile{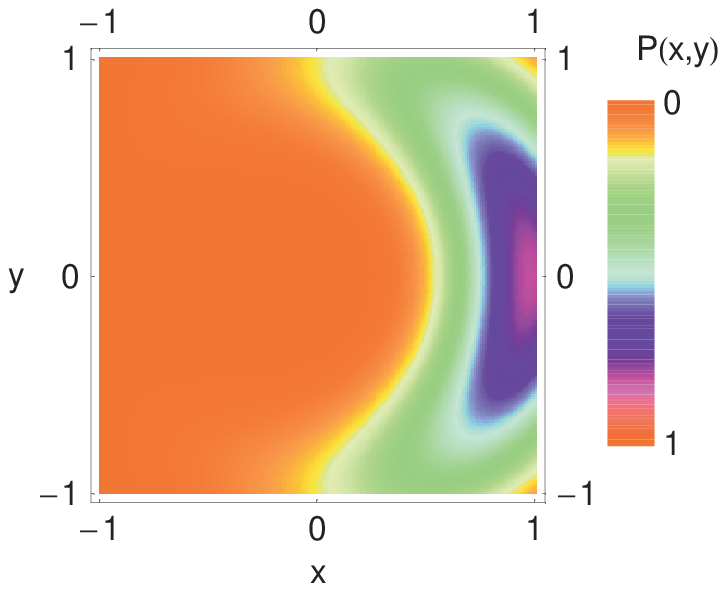}}
\end{center}
\caption{\label{fig4}Color density plot representing the two-dimensional probability distribution, $P(x,y)$, of the free end, for $L/L_p=2/3$, obtained from the analytic approximation (Eq.(\ref{complete})). The longitudinal position, $x$, and the transverse position, $y$, have been rescaled to $x/L$ and $y/L$, repsectively.}
\end{figure}

Besides providing simple approximations for the probability distributions of transverse and longitudinal displacements, the effective-medium approach can also be used in order to provide an approximation for the complete distribution function, $G(x,y,\theta)$:
\begin{eqnarray}
\label{complete}
& &G(x,y,\theta)=\sqrt{\frac{18 L_p^3}{{\pi}^3L^7}}\exp\Big\{-\frac{L_p\theta^2}{2L}\Big\}\nonumber\\
& &\times\exp\Big\{-\frac{6L_p}{L^3}[(x-\cos\theta L)^2+(y-\sin\theta L)^2]\Big\}
\end{eqnarray}
Integrating out the angle, we obtain the probability distribution function for the two-dimensional position of the free end, $P(x,y)$. This distribution, for $L/L_p=2/3$, is shown in the density plot of Fig. \ref{fig4}. Comparing it with Fig. 2a of Ref. \cite{Lat}, one can see that the effective-velocity approximation qualitatively agrees with the simulation, and the results for the position of the crest obtained from the the two approaches are even quantitatively close. The quantitative agreement improves as the polymer gets stiffer.


Out of mathematical curiosity (and not only!), one may want to explore a modified version of the diffusion-convection problem described by Eq. (\ref{wlceq2}). The new problem contains an extra, dimensionless parameter $\epsilon$ as the amplitude of the convective velocity:
\begin{eqnarray}
\label{multi}
\Big[\frac{\partial}{\partial L}+\epsilon\sin\theta \frac{\partial}{\partial y}- \frac{1}{L_p}\frac{\partial^2}{\partial \theta^2}\Big]G_y[y(L), \theta(L)]=0.
\end{eqnarray}
While Eq. (\ref{wlceq2}) has only one scale, the parameter $\epsilon$ in Eq. (\ref{multi}) causes a separation of scales between the diffusion process which acts on the time scale $\tau=L$ and the convection process which acts on the time scale $T=L /\epsilon$. A rigorous multiscale-perturbation-theory \cite{Hinch} analysis of this problem is deferred to future work. The implementation of the simple effective-medium approach, however, yields quite interesting results. We now consider the effective-velocity to be equal to $\langle v \rangle=\epsilon \sin \theta$, and we look at the dependence of $P(y)$ on $\epsilon$. It turns out that, as $\epsilon$ decreases, the width of the $L/L_p$-interval which exhibits bimodality shrinks, and, at a ``critical'' value $\epsilon_c\approx 0.66$, $P(y)$ becomes unimodal {\it for any} $L/L_p$. In the context of the wormlike chain, softening the relative strength of ``convection'' can be interpreted as relaxing the local inextensibility constraint. We leave the disappearence of bimodality with the softening of the local extensibility of the polymer as a conjecture, which will be elucidated in a future work.

Summarizing, in this paper, we used the analogy between the wormlike chain and a diffusion-convection system, and we applied an effective-medium approach to analytically account for some unexpected features in the distribution of the semiflexible polymer conformations. It is remarkable that this very simple approximation has been so successful for three qualitatively different types of convective flow: random flow \cite{Red1}, power-law shear flow \cite{Red2}, and - in our case - periodic flow.

\bigskip

P.B. thanks R. Metzler for interesting discussions.


\begin{thebibliography}{}
\bibitem{ftnt}
We should mention that the same equation is satisfied by any point along the polymer contour.
\end{thebibliography}
\end{document}